\documentclass[conference,onecolumn,draftcls]{IEEEtran}

\usepackage{amssymb,amsfonts}
\usepackage{algorithmic}
\usepackage{graphicx}
\usepackage{textcomp}
\usepackage{xcolor}

\usepackage{algorithm}
\usepackage{amsthm}
\usepackage{color}
\usepackage{stfloats}

\usepackage[utf8]{inputenc}
\usepackage[T1]{fontenc}
\usepackage{url}
\usepackage{ifthen}
\usepackage{cite}
\usepackage[cmex10]{amsmath} % Use the [cmex10] option to ensure complicance
\begin{document}
\title{Short Regular Girth-8 QC-LDPC Codes From Exponent Matrices with Vertical Symmetry}

% %%% Single author, or several authors with same affiliation:
% \author{%
%  \IEEEauthorblockN{Andrew R.~Barron}
%  \IEEEauthorblockA{Department of Statistics and Data Science\\
%                    Yale University\\
%                    New Haven, CT, USA\\
%                    Email: andrew.barron@yale.edu}
% }

%% Many authors with many affiliations:
 \author{%
   \IEEEauthorblockN{Guohua Zhang\IEEEauthorrefmark{1},
                     Aijing Sun\IEEEauthorrefmark{1},
                     Ling Liu\IEEEauthorrefmark{2},
                     and Yi Fang\IEEEauthorrefmark{3}}
   \IEEEauthorblockA{\IEEEauthorrefmark{1}%
                    School of Communications and Information Engineering, Xi'an University of Posts and Telecommunications, \\Xi'an, China,
                    zhangghcast@163.com, sunaijing@xupt.edu.cn}
   \IEEEauthorblockA{\IEEEauthorrefmark{2}%
                     Guangzhou Institute of Technology, Xidian University, Guangzhou, China,
                     liuling@xidian.edu.cn}
   \IEEEauthorblockA{\IEEEauthorrefmark{3}%
                     School of Information Engineering, Guangdong University of Technology, Guangzhou, China,
                     fangyi@gdut.edu.cn}
 }

\maketitle

%%%%%%
%% Abstract:
%% If your paper is eligible for the student paper award, please add
%% the comment "THIS PAPER IS ELIGIBLE FOR THE STUDENT PAPER
%% AWARD." as a first line in the abstract.
%% For the final version of the accepted paper, please do not forget
%% to remove this comment!
%%

\begin{abstract}
   To address the challenge of constructing short girth-8 quasi-cyclic (QC) low-density parity-check (LDPC) codes, a novel construction framework based on vertical symmetry (VS) is proposed. Basic properties of the VS structure are presented. With the aid of these properties, existing explicit constructions for column weights from three to five which can be transformed into the VS structure are sorted out. Then two novel explicit constructions with the VS structure which guarantee short codes are presented for column weights of three and six. Moreover, an efficient search-based method is also proposed to find short codes with the VS structure. Compared with the state-of-the-art benchmarks, both the explicit constructions and the search-based method presented in this paper can provide shorter codes for most cases. Simulation results show that the new shorter codes can perform almost the same as or better than the longer existing counterparts. Thus, the new shorter codes can fit better with the low-latency requirement for modern communication systems.
\end{abstract}

\section{Introduction}
High-performance short codes have important applications in a variety of modern communication systems, including the future 6G system. A low-density parity-check (LDPC) code \cite{TSF01,F04,SYH23} with larger girth (being an even number at least four) typically provides a larger minimum distance and thus is likely to exhibit better performance \cite{KS20,KKS22,KS22,DF23,FSM23,FSM23b}.
A quasi-cyclic (QC) LDPC code is defined as the null space of a sparse parity-check matrix composed of circulants, and thus can be compactly described by its associated exponent matrix and the circulant size \cite{F04,ASP24,LXCB24}. For the girth of 6, the problem of constructing short QC-LDPC codes has been partially solved by a couple of existing methods. By contrast, how to construct short girth-8 QC-LDPC codes has been a long-standing challenge for channel coding community \cite{F04}. In recent years, both explicit constructions (i.e. without using search) \cite{ZD15,MG20,WZZZ22} and search-based methods \cite{TBS17,KD20,KS20,TB22} for girth-8 codes with short lengths (or equivalently, small circulant sizes) have attracted increasing attention. In general, the smallest circulant sizes in the literature as far as we know are given by earliest-sequence (ES) method \cite{VPI04} ($J=3$) and max-function method \cite{ZSW12} ($J=4$) for explicit constructions, and by horizontal symmetry (HS)\cite{TBS17} and integer-ring-sieve (IRS) methods \cite{TB22} for search-based ones.

In this paper, we first present a new structure regarding exponent matrices, which is called vertical symmetry (VS) structure. An exponent matrix with the VS structure can be defined by $\textbf{E}_{vs}=[\textbf{E}_U^T, \textbf{E}_D^T]^T$ if $J$ is even, and $[\textbf{0}^T,\textbf{E}_U^T, \textbf{E}_D^T]^T$ otherwise, where $\textbf{E}_D=-\textbf{E}_U$. Based on the VS structure, we then propose several new explicit and search-based methods, which are able to offer the current smallest circulant size for most cases of row weights. Moreover, simulations show that the proposed shorter codes perform as well as or even better than the existing benchmarks for longer lengths. Therefore, the novel short codes have potential to match better with low-latency requirement for modern communication systems.

%Each paper must be classified as ``eligible for student paper award''
%or ``not eligible for student paper award''. Papers that are eligible for the student paper award must also include the comment ``THIS PAPER IS ELIGIBLE FOR THE STUDENT PAPER AWARD." as a first line in the abstract.

\section{Explicit Constructions for Odd Column Weights}

This section presents a basic property for VS exponent matrices with odd column weights, and then provides some explicit constructions for $J=3$ and $J=5$.

In this paper, only weight-1 circulant (i.e. cyclic permutation matrix, CPM) is considered. A $P\times P$ circulant associated with the respective element $e$ within the $J\times L$ exponent matrix, is defined as a $P\times P$ identity matrix with all its rows being cyclically shifted to the right by $\texttt{mod}(e,P)$ positions. Consequently, the parity-check matrix of a $(J,L)$-regular QC-LDPC code is a $J\times L$ array of $P\times P$ circulants.

\emph{Lemma 1}: Let $J\geq 3$ be odd. Suppose that the exponent matrix $\textbf{E}=[\alpha_0,\alpha_1,\cdots,\alpha_{J-1}]^T[\beta_0,\beta_1,\cdots,\beta_{L-1}]$ satisfies the constraint $\alpha_i+\alpha_{J-1-i}=2\alpha_\frac{J-1}{2}$ for each $i$ in the range $0\leq i<\frac{J-1}{2}$. Then $\textbf{E}$ is equivalent to a VS exponent matrix: $\textbf{E}_{vs}=[0,\textbf{a},-\textbf{a}]^T[\beta_0,\beta_1,\cdots,\beta_{L-1}]$, where $\textbf{a}=[\alpha_\frac{J+1}{2}-\alpha_\frac{J-1}{2},\alpha_{\frac{J+1}{2}+1}-\alpha_\frac{J-1}{2},\cdots,\alpha_{J-1}-\alpha_\frac{J-1}{2}]$.

\emph{Proof}: See Appendix. \qed

Lemma 1 plays an important role in the following two subsections.
\subsection{VS exponent matrices for $J=3$}
In this subsection, a type of VS exponent matrices for $J=3$ is proposed, which contains three specific cases (two derived from existing sequences and one from a new sequence).

Let $J=3$ and set $[\alpha_0,\alpha_1,\alpha_2]=[0,1,2]$. According to Lemma 1, $[0,1,2]^T[\beta_0,\beta_1,\cdots,\beta_{L-1}]$ has an equivalent VS exponent matrix $\textbf{E}_{vs}=[0,1,-1]^T[\beta_0,\beta_1,\cdots,\beta_{L-1}]$.

\emph{Theorem 1}: Let $\{\beta_0,\beta_1,\cdots,\beta_{L-1}\}$ be a sequence composed of $L$ distinct (modulo $P$) integers such that (1) $2\beta_i\neq 2\beta_j~(mod~P)$, where $0\leq i<j\leq L-1$; and (2) $2\beta_k\neq \beta_i+\beta_j~(mod~P)$, where $i$, $j$ and $k$ are different integers from $0$ to $L-1$. Then $[0,1,-1]^T[\beta_0,\beta_1,\cdots,\beta_{L-1}]$ is a VS exponent matrix, which yields a girth-8 QC-LDPC code for the circulant size $P$.

\emph{Proof}: See Appendix. \qed

In what follows, we consider three sequences which satisfy the inequality in Theorem 1. The first two sequences are existing ones in the literature, and the third is novel.

\emph{Definition 1}\cite{ZHFW19}\cite{ZW22}: If $\{s_0,s_1,\cdots,s_{L-1}\}$ is a sequence of integers such that all pair-wise sums $(s_i+s_j, 0\leq i\leq j\leq L-1)$ are distinct modulo $P$, then it is called a Sidon sequence over $\textbf{Z}_P$, the integer ring of modulo $P$.

\emph{Corollary 1}: Let $\{\beta_0,\beta_1,\cdots,\beta_{L-1}\}$ be a Sidon sequence over $\textbf{Z}_P$. Then $[0,1,-1]^T[\beta_0,\beta_1,\cdots,\beta_{L-1}]$ is a VS exponent matrix for a girth-8 QC-LDPC code with the circulant size $P$.

Since a Sidon sequence with $L$ elements can possibly exist only if $P\geq L(L-1)+1$, VS exponent matrices provided by Corollary 1 generally lead to large circulant sizes. To obtain short girth-8 codes with the VS structure, other sequences need to be examined. The earliest sequence (ES) belongs to such sequences allowing for girth-8 codes with short circulant sizes.

\emph{Definition 2 (ES)}\cite{VPI04}: If $n$ is even, set $s_{el}(n)=3\cdot s_{el}(n/2)$; otherwise, set $s_{el}(n)=s_{el}(n-1)+1$, where $s_{el}(0)=0$.

\emph{Corollary 2}: Let $\{\beta_0,\beta_1,\cdots,\beta_{L-1}\}$ consist of the first $L$ entries of the ES. Then $[0,1,-1]^T[\beta_0,\beta_1,\cdots,\beta_{L-1}]$ is a VS exponent matrix for a girth-8 QC-LDPC code with the circulant size $P=2\cdot s_{el}(L-1)+1$.

Generally speaking, as far as explicit constructions for column weight of three are concerned, the ES has held the best record in providing the shortest circulant size (i.e. $P$ in Corollary 2) since its invention in 2004. A thought-provoking question, therefore, is whether there exists a way to go beyond the unrivaled ES. In our search for exponent matrices by using a Two-Direction column-scanning greedy strategy, a magical sequence (called TD for short) is discovered. Through careful analysis, we realize that the TD is inextricably linked with the ES, and can be compactly defined via the latter.

\emph{Definition 3 (TD)}: Let $s_{td}(n)=(-1)^{n+1}[6\cdot s_{el}(\lfloor\frac{n}{4}\rfloor)+\texttt{mod}(n,4)]$ for each $n\geq 0$.

Let $\{\beta_0,\beta_1,\cdots,\beta_{L-1}\}$ be a sequence consisting of the first $L$ entries of a TD sequence, i.e., $\beta_i=s_{td}(i)$ for $0\leq i\leq L-1$. Define the value of $P(L)$ by an iterative manner: $P(L)=3\cdot P(L/2)$ for an even $L\geq 4$ and $P(L)=3\cdot P(\frac{L+1}{2})+\texttt{mod}(L,4)-5$ for an odd $L\geq 3$, where $P(2)=3$.

\begin{figure}[h]
  \centering
  \includegraphics[width=15.0 cm,bb=0 0 400 300]{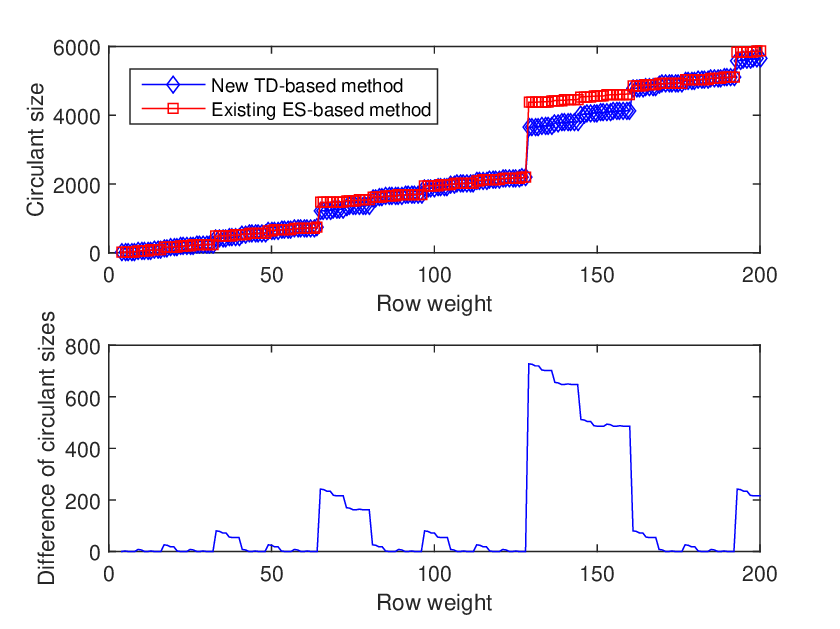}
  \caption{Circulant size comparison of two explicit VS methods for $J=3$: the proposed TD-based method and existing ES-based method.}
  \label{fig1}
\end{figure}

\emph{Conjecture 1}: Let $\{\beta_0,\beta_1,\cdots,\beta_{L-1}\}$ be a sequence consisting of the first $L$ entries of the TD. For the circulant size $P(L)$, the VS exponent matrix $[0,1,-1]^T[\beta_0,\beta_1,\cdots,\beta_{L-1}]$ generates a girth-8 QC-LDPC code.

The validity of Conjecture 1 has been verified for each $L$ in the range of $3\leq L\leq 500$. As shown in Fig. 1, for the vast majority of cases the TD offers a smaller circulant size than ES, and for the rest cases the former provides the same circulant size as ES.

Besides the single circulant size of $P(L)$, many circulant sizes larger than $2\cdot s_{el}(L-1)$ also allow girth-8 codes.

\emph{Conjecture 2}: The exponent matrix also leads to a $(3,L)$-regular girth-8 QC-LDPC code for any \emph{odd} circulant size $P\geq P_x$, where $P_x=2\cdot s_{el}(L-1)+1$.

For each $L$ in the range $3\leq L\leq 100$, Conjecture 2 has been verified for each odd circulant size $P$ in the range of $P_x\leq P<2\cdot P_x$. This indicates that for a general row weight, the new method not only provides the smallest circulant size to the best of our knowledge, in the sense of explicit constructions, but also allows for rather flexible circulant sizes.

\subsection{VS exponent matrices for $J\geq 5$}
Let $\beta_i=f_J(i)$ for $0\leq i\leq L-1$, where the function $f_J(i)$ is defined by the base-expansion method in Section IV of \cite{ZZ14}. Then  $\textbf{E}=[0,1,\cdots,J-1]^T[\beta_0,\beta_1,\cdots,\beta_{L-1}]$ corresponds to a girth-8 $(J,L)$-regular QC-LDPC code for each circulant size larger than $f_J(L-1)\cdot(J-1)$, according to the proof of Construction 5 in \cite{ZZ14}. Since $\alpha_i+\alpha_{J-1-i}=2\alpha_\frac{J-1}{2}=J-1$ for each $i$ in the range $0\leq i<\frac{J-1}{2}$, the constraint in Lemma 1 holds. Therefore, according to Lemma 1, $\textbf{E}$ has an equivalent VS form $\textbf{E}_{vs}=[\textbf{0},\textbf{a},-\textbf{a}]^T[\beta_0,\beta_1,\cdots,\beta_{L-1}]$, where $\textbf{a}=[1,2,\cdots,(J-1)/2]$.

Although VS exponent matrices with odd values of $J$ can be generated without search via the base-expansion method \cite{ZZ14} and Lemma 1, their circulant sizes are generally not small. As for an odd $J\geq 5$, we have not yet found an effective way to explicitly construct VS exponent matrices with the smallest possible circulant sizes, but this issue clearly deserves further investigation.

\section{Explicit Constructions for Even Column Weights}
This section presents a basic property for VS matrices with even column weights, and then provides some explicit constructions for $J=4$ and $J=6$.

\emph{Lemma 2}: Let $J\geq 4$ be even. If an exponent matrix $\textbf{E}=[\alpha_0,\alpha_1,\cdots,\alpha_{J-1}]^T[\beta_0,\beta_1,\cdots,\beta_{L-1}]$ satisfies $\alpha_i+\alpha_{J-1-i}=2x$ for each $i~(0\leq i\leq J/2-1)$, where $x$ is a constant, then $\textbf{E}$ is equivalent to a VS
exponent matrix $\textbf{E}_{vs}=[\textbf{a},-\textbf{a}]^T[\beta_0,\beta_1,\cdots,\beta_{L-1}]$, where $\textbf{a}=[\alpha_\frac{J}{2}-x,\cdots,\alpha_{J-1}-x]$.

\emph{Proof}: Lemma 2 can be similarly proved as Lemma 1. \qed

Lemma 2 has a crucial role in the following two subsections.

\subsection{VS exponent matrices for $J=4$}
In this subsection, a type of VS exponent matrices for $J=4$ is proposed, from which three specific cases are generated by combining Lemma 2 and certain existing constructions (based on greatest common divisor (GCD) \cite{ZSW13,ZZ14,ZFL19,ZLLF23}, disjoint difference sets (DDS) \cite{ZNHY23} and max function \cite{ZSW12}).

Let $e(i,r)$ be the entry in the $i$-th row and $r$-th column of a $J\times L$ matrix $\textbf{E}$, where $0\leq i\leq J-1$ and $0\leq r\leq L-1$. Then $\textbf{E}$ can be expressed as $\textbf{E}=[e(i,r)]$ for brevity.

\emph{Lemma 3}: Suppose that $\textbf{E}=[e(i,r)]$ is a $4\times L$ exponent matrix satisfying $e(0,r)=0$ and $e(3,r)=e(1,r)+e(2,r)$ for each $r~(0\leq r\leq L-1)$. Then $\textbf{E}$ can be equivalently transformed into a VS matrix
$\textbf{E}_{vs}$:
\begin{equation}
\left[
  \begin{array}{cccc}
a(0) & a(1) & \cdots& a(L-1)\\
b(0) & b(1) & \cdots& b(L-1)\\
-a(0) & -a(1) & \cdots& -a(L-1)\\
-b(0) & -b(1) & \cdots& -b(L-1)\\
\end{array}
\right]
\end{equation}
for the following two cases: (i) $e(1,r)+e(2,r)$ being even for each $r~(0\leq r\leq L-1)$: in this case, set $a(r)=-[e(1,r)+e(2,r)]/2$ and $b(r)=[e(1,r)-e(2,r)]/2$. (ii) $P$ being odd: in this case, if $e(1,r)+e(2,r)$ is even, set $a(r)=-[e(1,r)+e(2,r)]/2$ and $b(r)=[e(1,r)-e(2,r)]/2$; otherwise, set $a(r)=-[P+e(1,r)+e(2,r)]/2$ and $b(r)=[e(1,r)-e(2,r)-P]/2$.

\emph{Proof}: See Appendix. \qed

The case 1 in Lemma 3 is illustrated by two examples. The first one is an explicit construction based on GCD, while the second is based on random search.

\emph{Example 1 (GCD)}: Let $J=4$ and $L$ be odd. Set $\textbf{E}=[0,1,L,L+1]^T[0,1,\cdots,L-1]$ \cite{ZSW13}. Then its VS form is $\textbf{E}_{vs}=[\textbf{a},-\textbf{a}]^T[0,1,\cdots,L-1]$, where $\textbf{a}=[(L-1)/2,(L+1)/2]$. The VS matrix corresponds to a $(4,L)$-regular girth-8 QC-LDPC code for any circulant size $P\geq L^2$.

\emph{Example 2}: Set $P=38$ and $L=8$. Generate the exponent matrix $\textbf{E}$ according to Lemma 3, where the middle two rows are $[e(1,0),e(1,1),\cdots,e(1,L-1)]=[7,10,20,11,29,2,35,12]$ and $[e(2,0),e(2,1),\cdots,e(2,L-1)]=[1,10,22,3,15,16,19,28]$, respectively. Then, thanks to Lemma 3 (case 1), it is easily seen that $\textbf{E}$ is equivalent to the following VS exponent matrix:  \begin{equation}
\left[
  \begin{array}{cccccccc}
    34 &  28  & 17 &  31  & 16 &  29 &  11 &  18\\
     3 &   0  & 37 &   4  &  7 &  31 &   8 &  30\\
   -34 & -28  &-17 & -31  &-16 & -29 & -11 & -18\\
    -3 &   0  &-37 &  -4  & -7 & -31 &  -8 & -30\\
\end{array}
\right],
\end{equation}
which corresponds to a $(4,8)$-regular girth-8 code with the circulant size of 38. The length of this code is shorter than the shortest one \cite{TBS17} in the literature as far as we know.

To illustrate the case 2 in Lemma 3, we first use an example based on DDS \cite{ZNHY23}, and then present an explicit construction which is a modification of the max-function method \cite{ZSW12}.

\emph{Example 3 (DDS)}: Set $P=131$ and $L=8$. Then the two sets, $\textbf{d}_1=[0,31,37,55,56,83,97,99]$ and $\textbf{d}_2=[0,12,17,21,47,50,57,70]$ form a $(P,L,2)$-DDS over $\textbf{Z}_P$. Due to Lemma 3 (case 2), a VS exponent matrix can be obtained as follows:
\begin{equation}
\left[
  \begin{array}{cccccccc}
     0 & 44 & 104 & 93 & 14 & 130 & 54 & 112\\
     0 & 75 & 10 & 17 & 70 & 82 & 20 & 80\\
     0 & -44 & -104 & -93 & -14 & -130 & -54 & -112\\
     0 & -75 & -10 & -17 & -70 & -82 & -20 & -80\\
\end{array}
\right],
\end{equation}
which corresponds to a $(4,8)$-regular girth-8 code with the circulant size of 131.

Since a $(P,L,2)$-DDS over $\textbf{Z}_P$ possibly exists only if $P\geq 2L(L-1)+1$ \cite{ZNHY23}, the circulant size ($\geq 2L(L-1)+1$) in Example 3 is much larger than that ($\geq L^2$) in Example~1. Circulant sizes noticeably smaller than that in Example~1, are redolent of our earlier construction for $J=4$ based on max function \cite{ZSW12}, which is considered as a scheme able to provide by far the smallest circulant sizes in an explicit manner \cite{TBS17}. The max-function method can be slightly modified to attain VS girth-8 codes for each odd circulant size larger than approximately $3L^2/4$, where $L$ can be arbitrarily chosen.

\emph{Corollary 3 (max-function)}: Let $\textbf{E}=[e(i,r)]$ be a $4\times L$ exponent matrix. Set $e(0,r)=0$, $e(1,r)=r$ and $e(3,r)=e(2,r)+e(1,r)$ for $0\leq r\leq L-1$, where $e(2,0)=0$ and $e(2,r+1)=e(2,r)+\texttt{max}(r+2,L-r)$ for $0\leq r\leq L-2$. For each odd $P\geq \lceil 3\cdot L^2/4\rceil+L-1$, there exists a VS exponent matrix $\textbf{E}_{vs}$ which corresponds to a $(4,L)$-regular girth-8 QC-LDPC code for the circulant size $P$.

\emph{Proof}: See Appendix. \qed

In summary, the circulant sizes available from Example 1 (GCD, $L$ odd) and Example 3 (DDS, $P$ odd) are approximately linear with $L^2$ and $2L^2$, respectively, while that from Corollary 3 (max-function, $P$ odd) is merely with $3L^2/4$. Clearly, explicit constructions suitable for even $P$ noticeably smaller than $L^2$, remain unsolved and deserve further research.
\subsection{VS exponent matrices for $J=6$}

In this subsection, new VS exponent matrices with small circulant sizes are proposed for $J=6$.

\emph{Theorem 2}: The VS exponent matrix $\textbf{E}_{vs}=[\textbf{a},-\textbf{a}]^T[0,1,\cdots,L-1]$ corresponds to a $(6,L)$-regular girth-8 QC-LDPC code for the circulant size $P$, where $\textbf{a}$ and $P$ can be separately determined for four cases: (i) if $\texttt{mod}(L,6)\in\{0,2\}$, set $\textbf{a}=[2,L+1,L+3]$ and $P=(L+2)^2+3$; % (need proof);
(ii) if $\texttt{mod}(L,6)\in\{1,3\}$, choose $\textbf{a}=[2,L,L+2]$ and $P=(L+1)^2+3$; %(see TIT-submission)
(iii) if $\texttt{mod}(L,6)=4$, set $\textbf{a}=[2,L+3,L+5]$ and $P=(L+1)(L+5)$; % (need proof);
(iv) if $\texttt{mod}(L,6)=5$, choose $\textbf{a}=[2,L+2,L+4]$ and $P=L(L+4)$. %(see TIT-submission)

\emph{Proof}: See Appendix. \qed

From Fig. 2, it is clear that the circulant sizes offered by the new girth-8 codes in Theorem 3 are much shorter than existing benchmarks for almost all values of row weight.

\begin{figure}[h]
  \centering
  \includegraphics[width=15 cm,bb=0 0 400 300]{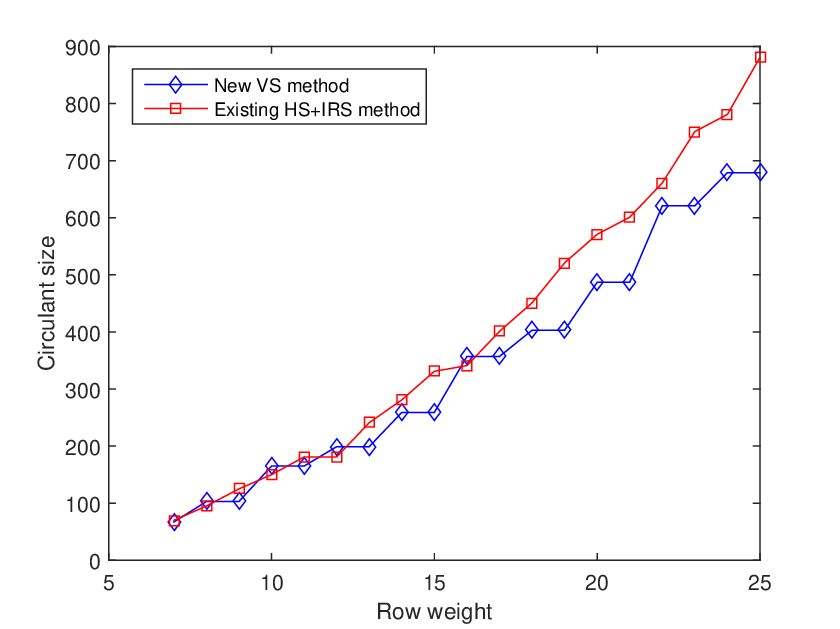}
  \caption{Circulant size comparison for $J=6$: the proposed explicit VS method and existing random (HS+IRS) method.}
  \label{fig3}
\end{figure}

\section{Search-based method}
In Section II and III, VS exponent matrices are proposed via pure formulae without search procedure. In this section, we consider how to obtain VS exponent matrices with the aid of simple but efficient search procedure. To be specific, we only consider a type of VS exponent matrices, the upper part of which can be expressed as the product of two sequences: $\{\alpha_0,\alpha_1,\cdots\,\alpha_{\lfloor\frac{J-2}{2}\rfloor}\}$ and $\{\beta^0,\beta^1,\cdots,\beta^{L-1}\}$.
Set $J_0=\lfloor\frac{J-2}{2}\rfloor$. Define $\textbf{E}_U=[\alpha_i\cdot \beta^r]$ for $0\leq i\leq J_0$ and $0\leq r\leq L-1$.

\subsection{Governing equations for 4-cycles}
Let $r$ and $s$ be two indexes such that $0\leq r<s\leq L-1$.

(1) If a 4-cycle exists in the upper (or lower) part of the exponent matrix, then such a cycle can be represented by $(\alpha_i-\alpha_j)(\beta^r-\beta^s)=0~(mod~P)$,
where $0\leq i<j\leq J_0$.
(2) If a 4-cycle occurs cross the upper and lower parts of the exponent matrix, then this cycle can be denoted by $(\alpha_i+\alpha_j)(\beta^r-\beta^s)=0~(mod~P)$,
where $0\leq i\leq j\leq J_0$.
(3) For a 4-cycle in the all-zero row and the upper (or lower) part, it can be expressed as $\alpha_i(\beta^r-\beta^s)=0~(mod~P)$,
where $0\leq i\leq J_0$.

\subsection{Governing equations for 6-cycles}
Let $r$, $s$ and $t$ be different indexes in the range $0\leq r,s,t\leq L-1$.
(1) If a 6-cycle exists in the upper (or lower) part of the exponent matrix, then such a cycle can be represented by $(\alpha_i-\alpha_j)\cdot\beta^s+(\alpha_j-\alpha_k)\cdot\beta^r+(\alpha_k-\alpha_i)\cdot\beta^t =0~(mod~P)$, where $0\leq i<j<k\leq J_0$.

(2a) If a 6-cycle occurs in two rows in the upper part and one row in the lower part of the exponent matrix, then this cycle can be expressed by $(\alpha_i-\alpha_j)\cdot\beta^s+(\alpha_j+\alpha_k)\cdot\beta^r-(\alpha_k+\alpha_i)\cdot\beta^t =0~(mod~P)$
where $0\leq i<j\leq J_0$ and $0\leq k\leq J_0$.
(2b) If a 6-cycle exists in one row in the upper part and two rows in the lower part of the exponent matrix, then this cycle can be denoted by $(\alpha_i+\alpha_j)\cdot\beta^s+(\alpha_k-\alpha_j)\cdot\beta^r-(\alpha_k+\alpha_i)\cdot\beta^t =0~(mod~P)$
where $0\leq i\leq J_0$ and $0\leq j<k\leq J_0$.

(3a) If a 6-cycle occurs in the all-zero row and the upper (or lower) part of the exponent matrix, then such a cycle can be represented by $\alpha_i(\beta^r-\beta^s)+\alpha_j(\beta^t-\beta^r)=0~(mod~P)$,
where $0\leq i<j\leq J_0$.
(3b) If a 6-cycle exists in the all-zero row, the upper part, and the lower part of the exponent matrix, then this cycle can be expressed as $\alpha_i(\beta^r-\beta^s)-\alpha_j(\beta^t-\beta^r)=0~(mod~P)$,
where $0\leq i,j\leq J_0$.

\subsection{Search algorithm}
To reduce search space, $\alpha_0$ is set to 1. Consequently, huge search space due to $L\lfloor J\rfloor/2$ values is significantly reduced to very small space merely involving $\lfloor J\rfloor/2$ values. For example, when $J=4$ and $J=5$, it suffices to search for only two values ($\alpha_1$ and $\beta$) to generate a $(J,L)$-regular QC-LDPC code without cycles of four and six. For another example, when $J=6$ and $J=7$, only three values ($\alpha_1$, $\alpha_2$ and $\beta$) are enough to define the new QC-LDPC codes. The results of this search method (by avoiding all 4-cycles and 6-cycles described in Subsection B) are listed in Tables II$\sim$IV. Compared with the state-of-the-art benchmarks (HS and IRS methods), the new search method can provide shorter codes for most values of $L$ (especially for $J=5$ and $J=6$).

 \begin{table}[h]
   % increase table row spacing, adjust to taste
%   \renewcommand{\arraystretch}{1.3}
   \caption{Circulant size comparison for $J=4$: new search method and the state-of-the-art search method \cite{TBS17} (NA: not available).}
   \label{t1}
   \centering
\begin{tabular}{c|cc|c}
\hline
$L$ & $P_{new}$ & $[\alpha_1, \beta]$ & $P_{HS}$\cite{TBS17}\\\hline
5 & 29 & [12,5] & 23\\
6 & 37 & [3,11] & 24\\
7 & 43 & [12,4] & 30\\
8 & 53 & [23,3] & 39\\
9 & 61 & [24,3] & 48\\
10 & 61 & [24,3] & 57\\
11 & 89 & [4,2] & 67\\
12 & 91 & [31,19] & 80\\
13 & 131 & [17,39] & 98\\
14 & 137 & [37,16] & 112\\
15 & 137 & [37,16] & 130\\
\textbf{16} & 137 & [37,16] & 150\\
\textbf{17} & 137 & [37,16] & 170\\
\textbf{18} & 181 & [72,101] & 190\\
\textbf{19} & 199 & [74,124] & 205\\
\textbf{20} & 199 & [74,124] & 220\\
21 & 199 & [74,124] & NA\\
22 & 277 & [4,16] & NA\\
23 & 277 & [4,16] & NA\\
24 & 313 & [25,19] & NA\\
\textbf{25} & 313 & [25,19] & 350\\
\hline
\end{tabular}
\end{table}

 \begin{table}[h]
   % increase table row spacing, adjust to taste
%   \renewcommand{\arraystretch}{1.3}
   \caption{Circulant size comparison for $J=5$: new search method and and the state-of-the-art search method \cite{TBS17}.}
   \label{t2}
   \centering
\begin{tabular}{c|cc|c}
\hline
$L$ & $P_{new}$ & $[\alpha_1, \beta]$ & $P_{HS}$\cite{TBS17}\\\hline
6 & 49 & [6,19] & 35\\
7 & 71 & [2,20] & 53\\
8 & 73 & [27,2] & 64\\
\textbf{9} & 73 & [27,2] & 81\\
\textbf{10} & 89 & [34,2] & 99\\
\textbf{11} & 89 & [34,2] & 121\\
\textbf{12} & 137 & [37,9] & 142\\
13 & 185 & [43,113] & 165\\
14 & 211 & [28,12] & 190\\
\textbf{15} & 217 & [100,122] & 225\\
\textbf{16} & 233 & [89,3] & 250\\
\textbf{17} & 233 & [89,3] & 280\\
\textbf{18} & 233 & [89,5] & 320\\
\textbf{19} & 289 & [38,57] & 360\\
\textbf{20} & 289 & [38,57] & 400\\
21 & 313 & [25,19] & NA\\
22 & 313 & [25,19] & NA\\
23 & 313 & [25,19] & NA\\
24 & 313 & [25,19] & NA\\
\textbf{25} & 313 & [25,19] & 590\\
\hline
\end{tabular}
\end{table}

 \begin{table}[h]
   % increase table row spacing, adjust to taste
%   \renewcommand{\arraystretch}{1.3}
   \caption{Circulant size comparison for $J=6$: new search method and the state-of-the-art search methods \cite{TBS17,TB22}.}
   \label{t3}
   \centering
\begin{tabular}{c|cc|c|c}
\hline
$L$ & $P_{new}$ & $[\alpha_1, \alpha_2, \beta]$ & $P_{HS}$\cite{TBS17} & $P_{IRS}$\cite{TB22}\\\hline
7 & 97 & [35,36,43] & 70 & 101\\
8 & 109 & [8,37,16] & 95 & 121\\
\textbf{9} & 109 & [8,37,16] & 125 & 151\\
\textbf{10} & 143 & [11,12,28] & 150 & 181\\
\textbf{11} & 169 & [26,64,46] & 182 & 181\\
\textbf{12} & 169 & [65,77,19] & 218 & 181\\
13 & 271 & [47,110,30] & 254 & 241\\
14 & 289 & [16,68,5] & 296 & 281\\
\textbf{15} & 289 & [34,135,99] & 337 & 331\\
\textbf{16} & 289 & [34,135,99] & 380 & 341\\
\textbf{17} & 361 & [18,133,33] & 429 & 401\\
\textbf{18} & 361 & [38,75,33] & 478 & 451\\
\textbf{19} & 451 & [7,147,46] & 530 & 521\\
\textbf{20} & 451 & [7,147,46] & 584 & 571\\
21 & 529 & [22,46,28] & NA & 601\\
22 & 529 & [22,46,28] & NA & 661\\
23 & 599 & [35,267,18] & NA & 751\\
24 & 601 & [45,189,2] & NA & 781\\
25 & 601 & [45,189,2] & NA & 881\\
\hline
\end{tabular}
\end{table}

Moreover, the following example shows that by using a random search without the constraint of $\textbf{E}_U=[\alpha_i\cdot \beta^r]$, the VS structure we proposed can provide the current smallest circulant sizes for $J=4$ and small values of $L$.

\emph{Example 4}: Let $J=4$. Then the following two $\textbf{E}_U$'s lead to a VS girth-8 $(4,5)$-regular code with $P=21$ and a VS girth-8 $(4,7)$-regular code with $P=29$, respectively.

\begin{equation}\nonumber
%\textbf{E}_{BTS45}=
\left[
  \begin{array}{cccccccc}
     0 & 1 & 2 & 3 & 8\\
     0 & 8 & 3 & 7 & 10\\
\end{array}
\right],
\left[
  \begin{array}{cccccccc}
     0 & 1 & 2 & 4 & 9 & 25 & 26\\
     0 & 2 & 12 & 9 & 11 & 19 & 14\\
\end{array}
\right].
\end{equation}

\section{Performance simulations}
In this section, three new VS codes are compared with existing  benchmarks in terms of bit-error rate (BER) and block-error rate (BLER). Assume the BPSK modulation, AWGN channel and sum-product-algorithm (SPA) decoding. (1) for $J=3$ and $L=9$, the new TD-based VS code is compared with existing ES-based code; (2) for $J=6$ and $L=11$, the new code (generated by Theorem 2 (iv)) is compared with the two codes from IRS-search method \cite{TB22} and HS-search method \cite{TBS17}; (3) for $J=6$ and $L=12$, the new search-based code is compared with the code from IRS-search method \cite{TB22}. From Fig. 3$\sim$5, it is observed that the new shorter VS codes perform almost the same as or better than the longer counterparts. Also shown in Fig.~5 is the noticeably improved performance of a masked $(3,6)$-regular code, which is generated by combining the new VS code with a masking matrix (from $\textbf{g}_1=[1,0,0,0,1,1]$ and $\textbf{g}_2=[0,1,0,0,1,1]$ via the manner described in Example 2 (code 1) of \cite{ZFL19}).

\begin{figure}[h]
  \centering
  \includegraphics[width=15 cm,bb=0 0 400 300]{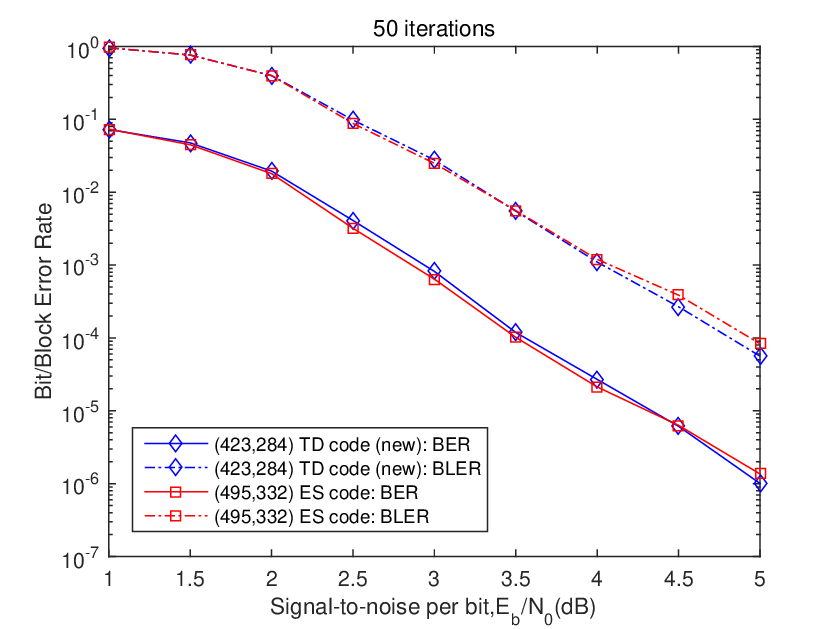}
  \caption{Performance comparison of $(3,9)$-regular girth-8 VS codes: derived from existing ES and new TD methods.}
  \label{fig2}
\end{figure}

\begin{figure}[h]
  \centering
  \includegraphics[width=15 cm,bb=0 0 400 300]{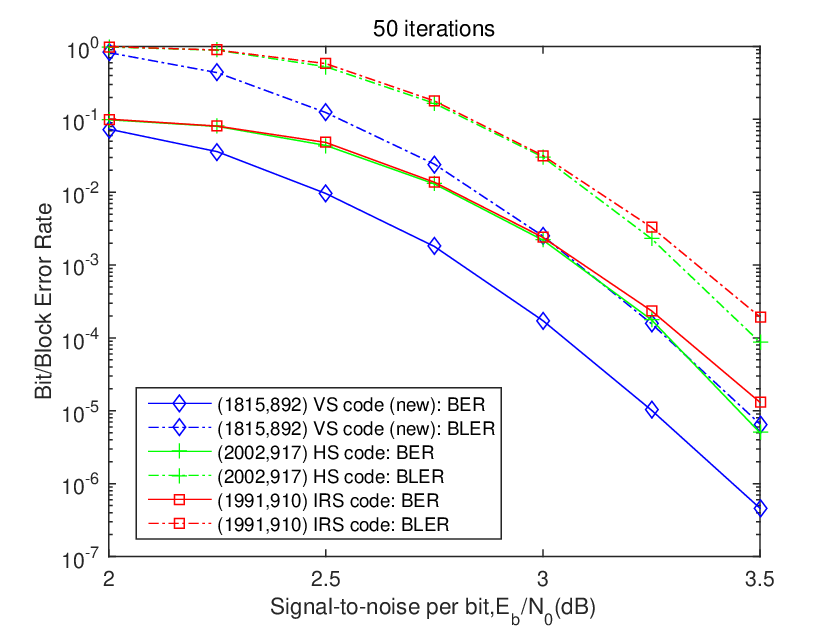}
  \caption{Performance comparison of $(6,11)$-regular girth-8 codes: new VS code ($L=11$ in Theorem 2 (iv)) and existing HS and IRS codes.}
  \label{fig4}
\end{figure}
%\vspace{-0.3cm}
\begin{figure}[h]
  \centering
  \includegraphics[width=15 cm,bb=0 0 400 300]{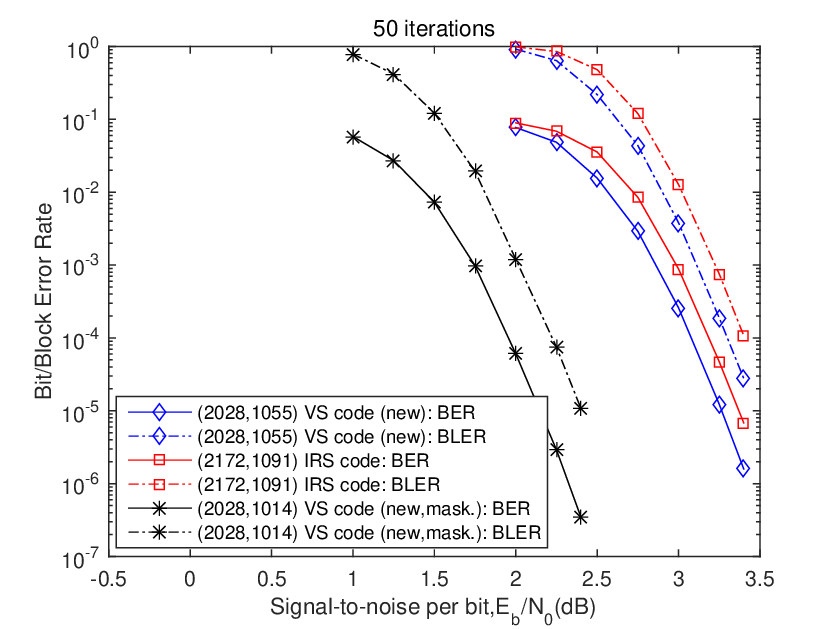}
  \caption{Performance comparison of $(6,12)$-regular girth-8 codes: new VS code ($L=12$ in Table \ref{t3}) and existing IRS code.}
  \label{fig5}
\end{figure}

\section{Conclusion}
A new structure called vertical symmetry (VS) for exponent matrices of QC-LDPC codes is proposed to construct short codes without cycles of length four and six. Properties, explicit constructions and search-based methods regarding the novel structure are investigated.
While the VS-based QC-LDPC codes have shorter lengths for most row weights, their performances are identical to or even superior to the existing benchmarks for longer lengths.

\section*{Acknowledgment}
This work was supported in part by the NSF of China under Grant 62322106, and the Guangdong Basic and Applied Basic Research Foundation under Grant 2022B1515020086.
%We are indebted to Michael Shell for maintaining and improving
%\texttt{IEEEtran.cls}.

%%%%%%
%% To balance the columns at the last page of the paper use this
%% command:
%%
%\enlargethispage{-1.2cm}
%%
%% If the balancing should occur in the middle of the references, use
%% the following trigger:
%%
%\IEEEtriggeratref{4}
%%
%% which triggers a \newpage (i.e., new column) just before the given
%% reference number. Note that you need to adapt this if you modify
%% the paper.  The "triggered" command can be changed if desired:
%%
%\IEEEtriggercmd{\enlargethispage{-20cm}}
%%
%%%%%%

%%%%%%
%% References:
%% We recommend the usage of BibTeX:
%%
%\bibliographystyle{IEEEtran}
%\bibliography{definitions,bibliofile}

\begin{thebibliography}{9}

\bibitem{TSF01}
R. M. Tanner, D. Sridhara, and T. E. Fuja, ``A class of group-structured LDPC codes,'' \emph{in Proc. Int. Symp. Commun. Theory and Applications}, pp. 1-5, Ambleside, U.K., Jul. 2001.

\bibitem{F04}
M. P. C. Fossorier, ``Quasi-cyclic low-density parity-check codes from circulant permutation matrices,'' \emph{IEEE Trans. Inf. Theory}, vol. 50, no. 8, pp. 1788-1793, Aug. 2004.

\bibitem{SYH23}
L. Song, S. Yu, and Q. Huang, ``Low-density parity-check codes: highway to channel capacity,'' \emph{China Commun.}, vol. 20, no. 2, pp. 235-256, Feb. 2023.

\bibitem{KS20}
I. Kim and H.-Y. Song, ``A simple construction for QC-LDPC codes of short lengths with girth at least 8,'' \emph{in 2020 International Conference on Information and Communication Technology Convergence (ICTC)}, pp. 1462-1465, Jeju, Korea(South), 21-23 Oct. 2020.

\bibitem{KKS22}
I. Kim, T. Kojima, and H.-Y. Song, ``Some short-length girth-8 QC-LDPC codes from primes of the form $t^2+1$,'' \emph{IEEE Commun. Lett.}, vol. 26, no. 6, pp. 1211-1215, Jun. 2022.

\bibitem{KS22}
I. Kim and H.-Y. Song, ``A construction for girth-8 QC-LDPC codes using Golomb rulers,'' \emph{Electron. Lett.}, vol. 58, no. 15, pp. 582-584, Jul. 2022.

\bibitem{FSM23}
A. G\'{o}mez-Fonseca, R. Smarandache, and D. G. M. Mitchell, ``A low complexity PEG-like algorithm to construct quasi-cyclic LDPC codes,'' \emph{in 12th International Symposium on Topics in Coding (ISTC)}, pp. 1-5, Brest, France, 04-08 Sept. 2023.

\bibitem{DF23}
H. Chimal-Dzul and A. G\'{o}mez-Fonseca, ``Using partial orthomorphisms to construct short quasi-cyclic LDPC codes with girth at least 6,'' \emph{in 2023 12th International Symposium on Topics in Coding (ISTC)}, pp. 1-5, Brest, France, 04-08 Sept. 2023.

\bibitem{FSM23b}
A. G\'{o}mez-Fonseca, R. Smarandache, and D. G. M. Mitchell, ``An efficient strategy to count cycles in the Tanner graph of quasi-cyclic LDPC codes,'' \emph{IEEE J. Sel. Area. Inf. Theory}, vol. 4, pp. 499-513, Sept. 2023.

\bibitem{ASP24}
F. Amirzade, M.-R. Sadeghi, and D. Panario, ``Construction of protograph-based LDPC codes with chordless short cycles,'' \emph{IEEE Trans. Inf. Theory}, vol. 70, no. 1, pp. 51-74, Jan. 2024.

\bibitem{LXCB24}
H. Li, H. Xu, C. Chen, and B. Bai, ``Efficient construction of quasi-cyclic LDPC codes with multiple lifting sizes,'' \emph{IEEE Commun. Lett.}, vol. 28, no. 4, pp. 754-758, Apr. 2024.

\bibitem{ZD15}
Y. Zhang and X. Da, ``Construction of girth-eight QC-LDPC codes from arithmetic progression sequence with large column weight,'' \emph{Electron. Lett.}, vol. 51, no. 16, pp. 1257-1259, Aug. 2015.

\bibitem{MG20}
M. Majdzade and M. Gholami, ``On the class of high-rate QC-LDPC codes with girth 8 from sequences satisfied in GCD condition,'' \emph{IEEE Commun. Lett.}, vol. 24, no. 7, pp. 1391-1394, Jul. 2020.

\IEEEtriggeratref{4}
\newpage

\bibitem{WZZZ22}
J. Wang, J. Zhang, Q. Zhou, and L. Zhang, ``Full-length row-multiplier QC-LDPC codes with girth eight and short
circulant sizes,'' \emph{IEEE Access}, vol. 11, pp. 22250-22265, Mar. 2023.

\bibitem{TBS17}
A. Tasdighi, A. H. Banihashemi, and M.-R. Sadeghi, ``Symmetrical constructions for regular girth-8 QC-LDPC codes,'' \emph{IEEE Trans. Commun.}, vol. 65, no. 1, pp. 14-22, Jan. 2017.

\bibitem{TB22}
A. Tasdighi and E. Boutillon, ``Integer ring sieve for constructing compact QC-LDPC codes with girths 8, 10, and 12,'' \emph{IEEE Trans. Inf. Theory}, vol. 68, no. 1, pp. 35-46, Jan. 2022.

\bibitem{KD20}
A. Kharin, A. Dryakhlov, E. Mirokhin, K. Zavertkin, A. Ovinnikov, and E. Likhobabin, ``An approach to the generation of regular QC-LDPC codes with girth 8,'' \emph{in Proc. 9th Medit. Conf. Embedded Comput. (MECO)}, pp. 1-4, Budva, Montenegro, 08-11 Jun. 2020.

\bibitem{VPI04}
B. Vasic, K. Pedagani, and M. Ivkovic, ``High-rate girth-eight low-density parity-check codes on rectangular integer lattices,'' \emph{IEEE Trans. Commun.}, vol. 52, no. 8, pp. 1248-1252, Aug. 2004.

\bibitem{ZSW12}
G. Zhang, R. Sun, and X. Wang, ``Explicit construction of girth-eight QC-LDPC codes and its application in CRT method,'' \emph{J. Commun.}, vol. 33, no. 3, pp. 171-176, Mar. 2012. (in Chinese).

\bibitem{ZHFW19}
G. Zhang, Y. Hu, Y. Fang, and J. Wang, ``Constructions of type-II QC-LDPC codes with girth eight from Sidon sequence,'' \emph{IEEE Trans. Commun.}, vol. 67, no. 6, pp. 3865-3878, Jun. 2019.

\bibitem{ZW22}
L. Zhang and J. Wang, ``Construction of QC-LDPC codes from Sidon sequence using permutation and segmentation,'' \emph{IEEE Commun. Lett.}, vol. 26, no. 8, pp. 1710-1714, Aug. 2022.

\bibitem{ZSW13}
G. Zhang, R. Sun, and X. Wang, ``Construction of girth-eight QC-LDPC codes from greatest common divisor,'' \emph{IEEE Commun. Lett.}, vol. 17, no. 2, pp. 369-372, Feb. 2013.

\bibitem{ZZ14}
J. Zhang and G. Zhang, ``Deterministic girth-eight QC-LDPC codes with large column weight,'' \emph{IEEE Commun. Lett.}, vol. 18, no. 4, pp. 656-659, Apr. 2014.

\bibitem{ZFL19}
G. Zhang, Y. Fang, and Y. Liu, ``Automatic verification of GCD constraint for construction of girth-eight QC-LDPC codes,'' \emph{IEEE Commun. Lett.}, vol. 23, no. 9, pp. 1453-1456, Sept. 2019.

\bibitem{ZLLF23}
G. Zhang, H. Liu, M. Lou, and Y. Fang, ``Constructions of girth-eight QC-LDPC codes with dual-diagonal structure based on GCD framework,'' \emph{in 2023 International Conference on Wireless Communications and Signal Processing (WCSP)}, pp. 269-274, Hangzhou, China, 02-04 Nov. 2023.

\bibitem{ZNHY23}
G. Zhang, M. Ni, Y. Hu, and Y. Yang, ``Quasi-cyclic LDPC codes with girth at least eight based on disjoint difference sets,'' \emph{IEEE Commun. Lett.}, vol. 27, no. 1, pp. 55-59, Jan. 2023.

\end{thebibliography}
%%
%% where we here have assumed the existence of the files
%% definitions.bib and bibliofile.bib.
%% BibTeX documentation can be obtained at:
%% http://www.ctan.org/tex-archive/biblio/bibtex/contrib/doc/
%%%%%%

%% Or you use manual references (pay attention to consistency and the
%% formatting style!):

\clearpage
\begin{appendices}
\section{Proof of Lemma 1}
\emph{Proof}: Clearly, after subtracting $\alpha_\frac{J-1}{2}\cdot\beta_i$ from each element within the $i$-th column of this matrix, the resultant matrix is equivalent to the original matrix. Perform this operation in turn for all columns, and the final matrix becomes $[\alpha_0-\alpha_\frac{J-1}{2},\alpha_1-\alpha_\frac{J-1}{2},\cdots,\alpha_{J-1}-\alpha_\frac{J-1}{2}]^T[\beta_0,\beta_1,\cdots,\beta_{L-1}]$
Since $\alpha_i+\alpha_{J-1-i}=2\alpha_\frac{J-1}{2}$, we have $\alpha_i-\alpha_\frac{J-1}{2}=-(\alpha_{J-1-i}-\alpha_\frac{J-1}{2})$. As a result, rearranging rows leads to the VS exponent matrix $\textbf{E}_{vs}=[0,\textbf{a},-\textbf{a}]^T[\beta_0,\beta_1,\cdots,\beta_{L-1}]$. \qed
 \section{Proof of Theorem 1}
\emph{Proof}: Thanks to Lemma 1, it suffices to prove that $[0,1,2]^T[\beta_0,\beta_1,\cdots,\beta_{L-1}]$ leads to a QC-LDPC code free of 4-cycles and 6-cycles for the circulant size $P$. First, consider 4-cycles.
Case (1): If there exist 4-cycles associated with the first two rows of the exponent matrix, then such cycles can be expressed as $(0-\beta_i)+(\beta_j-0)=0~(mod~P)$, where $0\leq i<j\leq L-1$. It reduces to $\beta_i=\beta_j~(mod~P)$, which is impossible due to Condition (i) in the theorem.
Case (2): If there are 4-cycles in the first and last rows of the exponent matrix, then they can be denoted by $(0-2\beta_i)+(2\beta_j-0)=0~(mod~P)$, where $0\leq i<j\leq L-1$. It becomes $2\beta_i=2\beta_j~(mod~P)$, which is impossible due to Condition (i) in the theorem.
Case (3): The nonexistence of 4-cycles associated with the last two rows of the exponent matrix can be similarly proved as in Case (1).

Next, consider 6-cycles. Let $i$, $j$ and $k$ be three different indexes from $0$ to $L-1$. If there exist 6-cycles in three rows of the exponent matrix, then without loss of generality they can be expressed as $(0-\beta_j+\beta_i-2\beta_i+2\beta_k-0)=0~(mod~P)$, equivalent to $2\beta_k=\beta_i+\beta_j~(mod~P)$. It is impossible owing to Condition (2) in the theorem.

Finally, the exponent matrix is not associated with 4-cycles but has a sub-matrix $[0,1,2]^T[\beta_i,\beta_j]$, where $0\leq i<j\leq L-1$. Therefore, 8-cycles exist \cite{ZZ14}.
\qed

\section{Proof of Lemma 3}
\emph{Proof}: (i) case 1: Subtracting $[e(1,r)+e(2,r)]/2$ from each element within the $r$-th column results in an equivalent exponent matrix. The $r$-th column becomes $[a(r),b(r),-b(r),-a(r)]^T$. Swapping the last two rows yields the final VS exponent matrix.  (ii) case 2: If $e(1,r)+e(2,r)$ is even, subtract $[e(1,r)+e(2,r)]/2$ from each element within the $r$-th column; otherwise, subtract $[P+e(1,r)+e(2,r)]/2$. The $r$-th column becomes $[a(r),b(r),-b(r),-a(r)]^T$. Swapping the last two rows yields the final VS exponent matrix. \qed

\section{Proof of Corollary 3}
\emph{Proof}: On the one hand, according to Theorem 1 in \cite{ZSW12}, the exponent matrix $\textbf{E}$ leads a girth-8 QC-LDPC code for each circulant size $P\geq \lceil 3L^2/4\rceil+L-1$. On the other hand, according to Lemma 3(ii), a VS exponent matrix $\textbf{E}_{vs}$ can be generated from  $\textbf{E}$ as long as the circulant size $P$ is odd. Therefore, $\textbf{E}_{vs}$ corresponds to a girth-8 QC-LDPC code for each odd circulant size $P\geq \lceil 3L^2/4\rceil+L-1$. \qed

\section{Proof of Theorem 2}

The following two properties and three transformations are useful in the proof of our new construction.

p.0: For a triple of integers $(0,a,b)$ such that $0<a<b$ and $b/\texttt{gcd}(b,a)\geq L$, the matrix $[0,a,b]^T\cdot[0,1,\cdots,L-1]$ has no 6-cycles for any circulant size $P\geq b(L-1)+1$.

p.1: For a triple of integers $(0,a,b)$ such that $\texttt{gcd}(a,c)=c$ and $\texttt{gcd}(b,c)=1$, the matrix $[0,a,b]^T\cdot[0,1,\cdots,L-1]$ has no 6-cycles for any circulant size $P$ satisfying $\texttt{gcd}(P,c)=c$.

Shifting (`S'): For any circulant size, the two exponent matrices, $[a_0,a_1,\cdots,a_i]^T\cdot[0,1,\cdots,L-1]$ and $[a_0-a_0,a_1-a_0,\cdots,a_i-a_0]^T\cdot[0,1,\cdots,L-1]$ have the same cycle distribution.

Reversion (`R') : For any circulant size,  the two exponent matrices, $[a_0,a_1,\cdots,a_i]^T\cdot[0,1,\cdots,L-1]$ and $[a_i-a_i,a_i-a_{i-1},\cdots,a_i-a_0]^T\cdot[0,1,\cdots,L-1]$ have the same cycle distribution.

Division (`D'): For any circulant size $P$ and any positive integer $d$ such that $\texttt{gcd}(d,P)=1$, the two exponent matrices, $[d\cdot a_0,d\cdot a_1,\cdots,d\cdot a_i]^T\cdot[0,1,\cdots,L-1]$ and $[a_0,a_1,\cdots,a_i]^T\cdot[0,1,\cdots,L-1]$ have the same cycle distribution. This transformation with the parameter $d$ is denoted by `/d'.

\emph{Proof}: We only prove the case (iii), i.e. $\texttt{mod}(L,6)=4$; the other cases can be similarly proved. It is equivalent to proving that $[0,2,L+3,L+7,2L+8,2L+10]^T[0,1,\cdots,L-1]$ corresponds to a girth-8 QC-LDPC code for the circulant size of $P=(L+1)(L+5)$. First, consider 4-cycles. There are 15 cases. The three cases,  expressed as $(0,2)$, $(0,L+3)$ and $(0,L+7)$, directly guarantee the absence of 4-cycles for $P=(L+5)(L+1)$. According to Transformation `D', the two cases, $(0,2L+8)$, $(0,2L+10)$, are turned into $(0,L+4)$ and $(0,L+5)$, respectively; By utilizing first Transformation `S' and then `D', the two cases, $(2,2L+8)$ and $(2,2L+10)$, are reduced to $(0,L+3)$ and $(0,L+4)$, respectively. Therefore, 4-cycles are also impossible for the four cases. For each of the rest 8 cases, it is reduced to the form $(0,x)$ via `S', where $2\leq x\leq L+7$. As a result, these 8 cases also exclude 4-cycles.

Now, we consider 6-cycles. There are 20 cases for 6-cycles, as listed in Table \ref{tc6}. The reason why 6-cycles for each case cannot exist is explained in the last column of Table \ref{tc6}.
For example, Case 12 means that the original triple $[2,L+3,2L+8]$ can be reduced to $[0,L+1,2L+6]$ via `S', and further to $[0,L+5,2L+6]$ via `R'.
Accordingly, the original matrix $[2,L+3,2L+8]^T\cdot[0,1,\cdots,L-1]$ is reduced to $[0,L+1,2L+6]^T\cdot[0,1,\cdots,L-1]$, and further to $[0,L+5,2L+6]^T\cdot[0,1,\cdots,L-1]$.
By choosing $a=L+5$, $b=2L+6$ and $c=L+5$, we now check whether $a$, $b$, $c$ and $P=(L+1)(L+5)$ satisfy all the conditions in Property p.1. Since $\texttt{mod}(L,6)=4$, it follows that $\texttt{gcd}(L+1,2)=1$ and hence $\texttt{gcd}(b,c)=\texttt{gcd}(L+1,4)=1$. Moreover, it is obvious that $\texttt{gcd}(P,c)=c$ and $\texttt{gcd}(a,c)=c$. Therefore, thanks to Property p.1, the matrix $[0,L+5,2L+6]^T\cdot[0,1,\cdots,L-1]$ are not associated with 6-cycles for the circulant size $P=(L+1)(L+5)$. Other 19 cases in this table can be similarly analyzed. \qed

\begin{table}[!htb]
%\scriptsize
\caption{Reasons for the nonexistence of 6-cycles associated with each combination of three rows within the exponent matrix. The notation $a\sim b$ means $\texttt{gcd}(a,b)=1$}
\centering
\begin{tabular}{l|lll}
\hline
$\#$ & Original triple & Reduced form & Reason\\\hline
1 & $[0,2,L+3]$ & - & $L$ even, p.0\\%p1 gcd
2 & $[0,2,L+7]$ & - & $L$ even, p.0\\
3 & $[0,2,2L+8]$ & $(/2)[0,1,L+4]$ & $P$ odd, p.0\\
4 & $[0,2,2L+10]$ &	$(/2)[0,1,L+5]$ & $P$ odd, p.0\\
5 & $[0,L+3,L+7]$ &	$(R)[0,4,L+7]$ & $L$ even, p.0\\
6 & $[0,L+3,2L+8]$ & $(R)[0,L+5,2L+8]$ & $L+1\sim2$, p.1\\
7 & $[0,L+3,2L+10]$	& - & $L+1\sim2$, p.1\\
8 & $[0,L+7,2L+8]$ & $(R)[0,L+1,2L+8]$ & $L+1\sim6$, p.1\\
9 & $[0,L+7,2L+10]$ & $(R)[0,L+3,2L+10]$ &	same as \#7\\\hline
10 & $[0,2L+8,2L+10]$ & $(R)[0,2,2L+10]$ & same as \#4\\
   &                  & $(/2)[0,1,L+5]$  &\\\hline
11 & $[2,L+3,L+7]$ & $(S)[0,L+1,L+5]$ & $L$ even, p.0\\
   &                  & $(R)[0,4,L+5]$ &\\\hline
12 & $[2,L+3,2L+8]$ & $(S)[0,L+1,2L+6]$ & $L+1\sim2$, p.1\\
   &                & $(R)[0,L+5,2L+6]$ &\\\hline
13 & $[2,L+3,2L+10]$ & $(S)[0,L+1,2L+8]$ & same as \#8\\
14 & $[2,L+7,2L+8]$ & $(S)[0,L+5,2L+6]$ & same as \#12\\
15 & $[2,L+7,2L+10]$ & $(S)[0,L+5,2L+8]$ & same as \#6\\\hline
16 & $[2,2L+8,2L+10]$ & $(S)[0,2L+6,2L+8]$ & same as \#3\\
   &                  & $(R)[0,2,2L+8]$ &\\
   &                  & $(/2)[0,1,L+4]$ &\\\hline
17 & $[L+3,L+7,2L+8]$ & $(S)[0,4,L+5]$ & same as \#11\\
18 & $[L+3,L+7,2L+10]$ & $(S)[0,4,L+7]$ & same as \#5\\\hline
19 & $[L+3,2L+8,2L+10]$ & $(S)[0,L+5,L+7]$ & same as \#2\\
   &                    & $(R)[0,2,L+7]$ &\\\hline
20 & $[L+7,2L+8,2L+10]$ & $(S)[0,L+1,L+3]$ & same as \#1\\
   &                    & $(R)[0,2,L+3]$ &\\
\hline
\end{tabular}\label{tc6}
\end{table}
\end{appendices}

\end{document}